# THE POLONNARUWA METEORITE:
# OXYGEN ISOTOPE, CRYSTALLINE AND BIOLOGICAL COMPOSITION

Jamie Wallis[2], Nori Miyake[1], Richard B. Hoover[1], Andrew Oldroyd[3], Daryl H. Wallis[1], Anil Samaranayake[4], K. Wickramarathne[4], M.K. Wallis[1], Carl H. Gibson[1,5] and N. C. Wickramasinghe[1†]

[1]*Buckingham Centre for Astrobiology, University of Buckingham, Buckingham, UK*
[2]*School of Mathematics, Cardiff University, Cardiff, UK*
[3]*School of Earth and Ocean Sciences, Cardiff University. Cardiff, UK*
[4]*Medical Research Institute, Colombo, Sri Lanka*
[5]*University of California San Diego, La Jolla, CA, 92093-0411, USA*

**Abstract:** Results of X-Ray Diffraction (XRD) analysis, Triple Oxygen Isotope analysis and Scanning Electron Microscopic (SEM) studies are presented for stone fragments recovered from the North Central Province of Sri Lanka following a witnessed fireball event on 29 December 2012. The existence of numerous nitrogen depleted highly carbonaceous fossilized biological structures fused into the rock matrix is inconsistent with recent terrestrial contamination. Oxygen isotope results compare well with those of CI and CI-like chondrites but are inconsistent with the fulgurite hypothesis.

**Keywords:** Polonnaruwa Meteorites, Carbonaceous Meteorites, Fulgurites, Diatoms, Comets, Panspermia

## 1. Introduction

A bright yellow fireball that turned green as it travelled across the sky was observed by several eyewitnesses in the North Central Province of Polonnaruwa, Sri Lanka at 18:30 PM on December 29, 2012. The green fireball was observed to disintegrate into sparkling fragments that fell to the ground near the villages of Aralaganwila and Dimbulagala and in a rice field (N 7° 52' 59.5" N; 81° 09' 15.7" E) near Dalukkane. The inferred NE to SW trajectory was determined from eyewitness observations and a distribution of stones recovered from a strewn field of >10 km. Police records indicate reports of low level burn injuries from immediate contact with the fallen stones and subsequent reports of a strong aroma. One woman was reported to have lost consciousness and was transported to the hospital after inhaling fumes from one of the stones. Witnesses reported that the newly fallen stones had a strong odour of tar or asphalt. Local police officials responded immediately by collecting samples and submitting them to the Medical Research Institute of the Ministry of Health in Colombo, Sri Lanka. Preliminary optical microscopic studies of deep interior samples extracted from the stones revealed the presence of a wide range of 10-40 µm-sized objects with a morphology characteristic of siliceous microalgae known as diatoms (Wickramasinghe et al., 2013a).

A sample of the recovered stones was then sent to us at Cardiff University, where we conducted studies of freshly cleaved interior surfaces using the Environmental Scanning Electron Microscope at the University's School of Earth and Ocean Sciences. These studies resulted in a number of images showing diatom frustules, some of which were clearly embedded in the rock matrix, thereby excluding the possibility of post-arrival contamination. Other structures of various shapes including large numbers of slender cylinders of lengths 5 - 10µm, and a few micrometers in diameter were seen to be distributed extensively throughout the sample (Wickramasinghe et al., 2013a). A separate sample was then sent to the United States, where it was investigated by one of us (RBH) using the Hitachi Field Emission Scanning Electron Microscope. This independent study on a different sample confirmed the presence of a range of diatom frustules, some of which were embedded in the rock matrix.

As a working hypothesis, we take the originating bolide to be of mass <100kg with a final explosive break-up occurring at about ~10km altitude, consistent with sightings of the fireball over tens of km and the dispersal of stones.

Initial examination of one stone indicated that it was uniquely inhomogeneous and poorly compacted with density less than 1g cm$^{-3}$ i.e. less than all known carbonaceous meteorites. The stone was predominantly black, though dark grey and grey regions could be visually detected together with evidence of a partial fusion crust. CHN analysis indicates an average carbon content of the order of 1-4% and combined with GC-MS spectra showing an abundance of high molecular weight organic compounds suggest tentative classification as a "carbonaceous chondrite ungrouped".

## 2. Samples

During the days and weeks that followed the initial collection of material in Polonnaruwa by local police officials, large quantities of stone artifacts were recovered from the rice fields in the vicinity of Aralaganwila and submitted for analysis. These included substantial quantities of stones recovered by the Department of Geology, University of Peradeniya as well as 628 separate fragments that were collected about a month later and subsequently submitted to us. The task of identifying broken fragments of meteorite from a mass of other stones was not easy. Of the 628 pieces examined, only three were clearly identified as possible meteorites after comparison with the original stones collected on the 29th December 2012. Of the samples rejected for further study, most were either, indigenous terrestrial stones, masonry composites or by-products of industrial processes. In addition to the above, a number of further samples were collected by us during a site visit on the 29th January 2013. These fragments were visibly different to the rejected material: they showed evidence of a partial, moderately thick fusion crust (0.25-0.5mm) with fractured surfaces exhibiting a highly porous composite structure characteristic of a carbonaceous chondrite, with fine grained olivine aggregates connected with mineral intergrowths. These samples were stored in sealed glass vials for later analysis.

---

[†] **Author correspondence** Professor N. C. Wickramasinghe Buckingham Centre for Astrobiology, mail: ncwick@gmail.com

For the purpose of this study, one of the original samples submitted by the Medical Research Institute in Colombo was selected for analysis. This fragment was portioned for interior section Scanning Electron Microscopy (SEM), oxygen isotope analysis, compositional analysis by X-Ray Diffraction (XRD) and elemental analysis by Inductively Coupled Plasma Optical Emission Spectroscopy (ICP-OES).

In addition to the portioned fragment, a sample of sand fulgurite was also analysed for comparative purposes together with a soil sample recovered from the fall site and a further sample of calcium-rich terrestrial rock selected for control purposes. Preliminary oxygen and carbon isotope analysis of the calcium rich control sample produced values of $\delta^{13}C = -9.716 \pm 0.02$ and $\delta^{18}O = -6.488 \pm 0.04$ which accord well with the values for soil carbonates determined by T.E. Cerling and J. Quade, 1993.

| Sample ID | Origin | Description |
|---|---|---|
| D159/001-01 | Dambulla, Sri Lanka | Soil Carbonate |
| F159/001-02 | Sahara Desert | Fulgurite |
| P159/001-03 | Aralaganwila, Sri Lanka | Witnessed Fall |
| P159/001-04 | Aralaganwila, Sri Lanka | Witnessed Fall |
| S159/001-05 | Aralaganwila, Sri Lanka | Soil Sample |

**Table 1** Details of samples analysed

### 3. X-Ray Diffraction Analysis

X-Ray Diffraction was carried out using a Philips PW1710 Automated Powder Diffractometer using Cu Kα radiation at 35kV and 40mA, between 2 and 70 °2θ at a scan speed of 0.04° 2θ/s. From the scans, phases were identified using PC-Identity software and from the peak areas, semi-quantitative analysis was performed and percentage of each phase calculated. Each of the four samples provided for analysis were powdered, mixed with a small amount of acetone and pipetted out onto glass slides. Sample preparation and analysis was conducted at the School of Earth Sciences at Cardiff University.

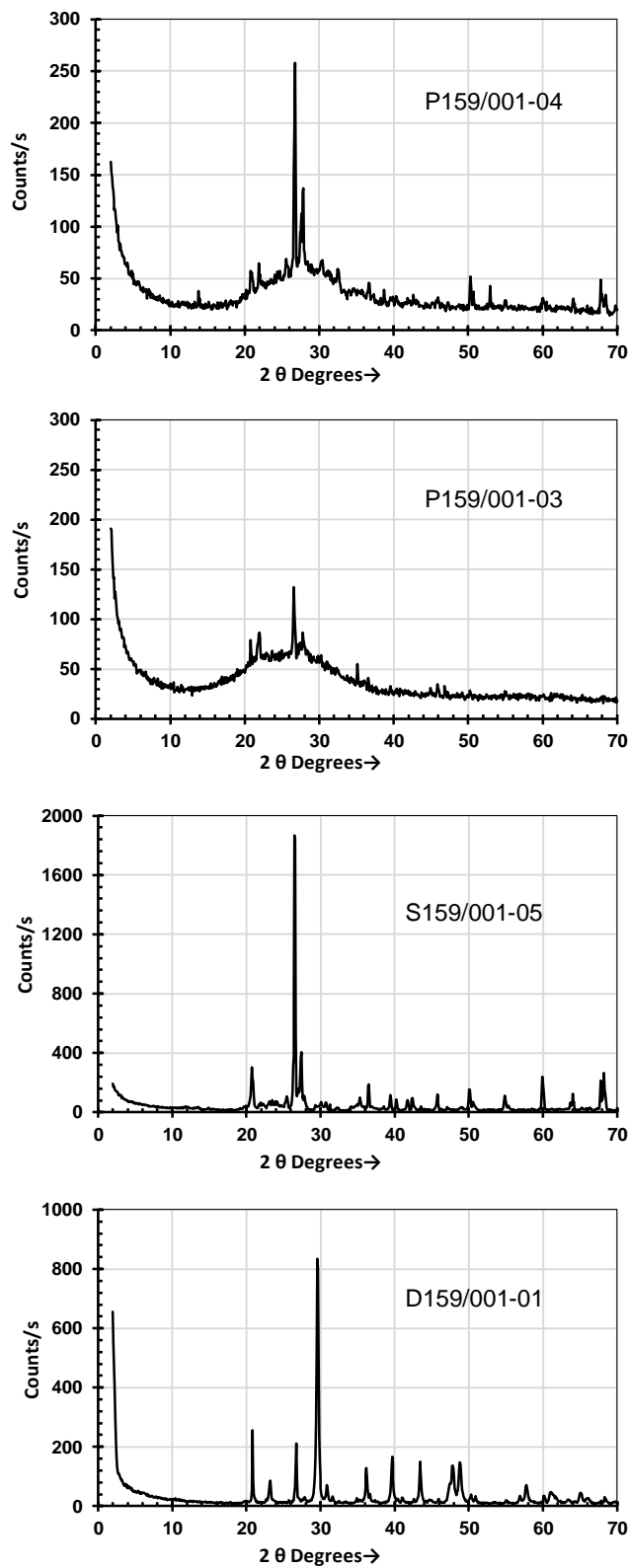

**Figure 1**. Shows X-Ray Diffraction patterns (counts/s against 2θ degrees) for samples P159/001-04, P159/001-03, S159/001-05 (soil) and D159/001-01 (control). The broad humps displayed in samples P159/001-04 and P159/001-03 are due to bulk composition of amorphous silica of (between 85-95%). The three crystaline phases detected are quartz (between 3-10%) together with anorthite (1-3%) and cristobaloite (approximately 2% in each sample).

X-Ray Diffraction patterns for the selected samples P159/001-04, P159/001-03, S159/001-05 and D159/001-01 (control) are represented in Figure 1. The broad hump displayed in P159/001-03 and P159/001-04 at ~18-36° $2\theta$ corresponds to a bulk composition of amorphous silica.

P159/001-03, P159/001-04 and S159/001-05 all display strong features at 20.8 $2\theta$ reported to be that of crystalline quartz. A strong feature at ~21.8 $2\theta$ in P159/001-04 was reported as $(NaK)AlSi_3O_8$ (anorthoclase) with further features in the 42-46 $2\theta$ range attributed to the presence of the silica polymorph cristobalite. P159/001-03 also reported the presence of $NaAlSi_3O_8$ (albite). Further weak but distinct peaks were noted at ~33.32 and 35.72 $2\theta$ in P159/001-04 with corresponding peaks at ~33.32, ~54.30 and ~35.72 $2\theta$ in P159/001-03 which is most likely due to the presence of $Fe_2O_3$ (hematite). Additional features at ~28.83 $2\theta$ in both P159/001-03 and P159/001-04 may be due to the presence of coesite. A weak feature at ~60.5 $2\theta$ suggests possible inclusions of stishovite but corresponding features at 30.18 and 45.76 $2\theta$ could not be determined with certainty and further work will be required to confirm this.

S159/001-05 (soil) reported the expected presence of approximately 75% quartz, 17% $KAlSi_3O_8$ (microcline), 4% $(NaK)AlSi_3O_8$, 3% $Al_2Si_2O_5(OH)_4$ (kaolinite) and 1% opal. Results reported for the control sample D159/001-01 were consisted with the available literature on soil carbonates comprising of 81% calcite, 13% quartz, 4% ankerite and 2% anorthite.

## 4. Triple Oxygen Isotope Analysis

Triple oxygen isotope analyses were conducted in the stable isotope laboratory at the University of Göttingen, Germany. Approximately 2mg of crushed sample was placed inside a Ni sample holder, evacuated overnight and heated to 70°C for 12h. An infrared (IR) laser (50W $CO_2$ laser, $\lambda$ = 10.6μm) was used to fluorinate the samples in purified $F_2$ gas under pressures of approximately 20 mbar. Sample $O_2$ was purified through the removal of excess $F_2$ by reaction with NaCl at 110°C to form NaF. $Cl_2$ gas was collected at a cold trap at -196°C. Sample oxygen was then collected at a 5Å molecular sieve at -196°C, expanded into a stainless steel capillary, transported with He carrier gas and re-trapped before release at 92°C through a 5Å molecular sieve GC-column of a Thermo Scientific GasBench II. The GC column was utilised to separate interfering $NF_3$, from $O_2$, as required for analysis of $^{17}O$ (Pack et al, 2008). The resulting purified sample $O_2$ was then expanded to the dual inlet system of a Thermo MAT 253 gas mass spectrometer. Results are presented in δ- and $\Delta^{17}O$-notation.

| | | |
|---|---|---|
| $\delta^{17}O$ | 8.978 | ± 0.050 |
| $\delta^{18}O$ | 17.816 | ± 0.100 |
| $\Delta^{17}O$ | -0.335 | ± 0.011 |

**Table 2**. Results of Oxygen Isotope Analysis

Data are reported relative to SMOW. Error bars are about ± 0.1 ‰ for $\delta^{17}O$ and $\delta^{18}O$ and about ± 0.01 ‰ or better for $\Delta^{17}O$. The reference gas was analysed relative to SMOW calibrated $O_2$ by E. Barkan (Hebrew University of Jerusalem). All δ values are permil unless otherwise stated.

Results are presented in δ-notation such that: $\delta^xO_{VSMOW}/1000 = [(^xO/^{16}O)_{sample} / (^xO/^{16}O)_{VSMOW}] - 1$. We use a slope of 0.5251 for the terrestrial fractionation line and linearized form of the δ-notation so that $\Delta^{17}O = \delta^{17}O - 0.5251 \delta^{18}O$. Error has been established through observation of the standard deviation in 290 single analyses of terrestrial rocks and minerals, carried out in 30 sessions at this stable isotope laboratory using the same equipment and is reported as +/- 0.06‰ (A. Gehler et al, 2011).

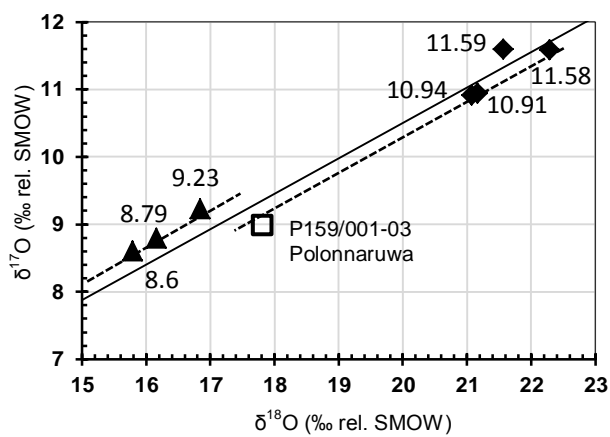

**Figure 2.** Oxygen isotope compositions of sample P159/001-04 and of CI chondrites: Alais (8.6), Ivuna (9.23) and Orgueil (8.79) together with CI-like chondrites (Meta-C) B-7904 (10.91), Y-82162 (11.59), Y-86720 (11.58) and Y-86789 (10.94). Data from Clayton and Mayeda, 1998.

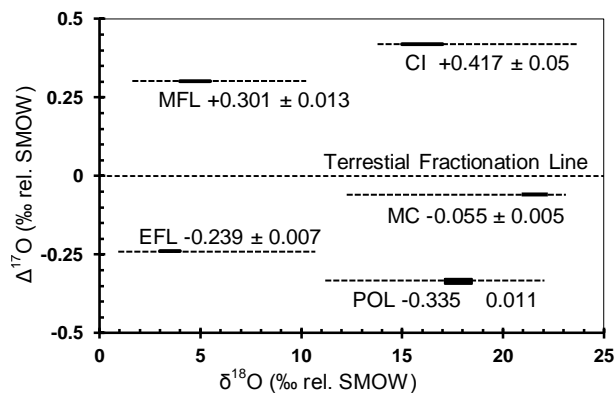

**Figure 3.** Oxygen isotope $\Delta^{17}O$ (‰ rel. SMOW) values of sample P159/001-04 (POL). Data for MFL: martian fractionation from Franchi et al. 1999. EFL: eucrite fractionation from Greenwood et al., 2005. CI: fractionation line for CI1 chondrites and MC: fractionation of Meta-C chondrites (B-9704 and Y-86789) from Clayton at al., 1998.

Results of analysis are displayed in Table 2, showing $\Delta^{17}O = -0.335$. Figure 2 shows a plot of $\delta^{17}O/\delta^{18}O$ values compared to those of known CI chondrites as reported by Clayton and Mayeda, 1998. It is noted that $\delta^{18}O$ values are relatively high but are within the range of the CI1 carbonaceous meteorites (Alais, Ivuna and Orgueil) and the CI-like and CM-like MCC (Metamorphosed Carbonaceous Chondrite) members on the new Meta-C group. Meteorites of the Meta-C group have been thermally metamorphosed on their parent body and have low water content and IR spectra indicating dehydration of their matrix phyllosilicates, very similar to our samples. Although the Meta-C meteorites somewhat resemble CI and CM carbonaceous chondrites, they are relatively depleted in Iron and Sulfur as compared to the CI and CM stones. The Meta-C meteorites show comparatively young Cosmic Ray Exposure (CRE) ages suggesting a near Earth asteroid as their common parent body. They have been recovered from Antarctica (Belgica-7904, Yamato 82162, Yamato 86720, and Yamato 86789) and the desert of Oman (Dhofar 225 and Dhofar 735 (Ikeda, 1992; (Ivanova et. al. 2002; Tameoka, 1990). The Meta-C meteorites show common Oxygen isotopic compositions that are entirely distinct from the Oxygen isotopic composition of CI1 and the CM1 and CM2 carbonaceous chondrites.

## 5. Scanning Electron Microscopy

Scanning Electron Microscopy (SEM) was conducted using the FEI (Phillips) XL30 FEG ESEM (Environmental Scanning Electron Microscope) FEG (Field Emission Gun) at the School of Earth Sciences at Cardiff University. The unit incorporates a secondary electron detector (SE), a back scatter electron detector (BSE) and a gaseous secondary electron detector (GSE). It also has an Oxford Instruments INCA ENERGY (EDX) x-ray analysis system.

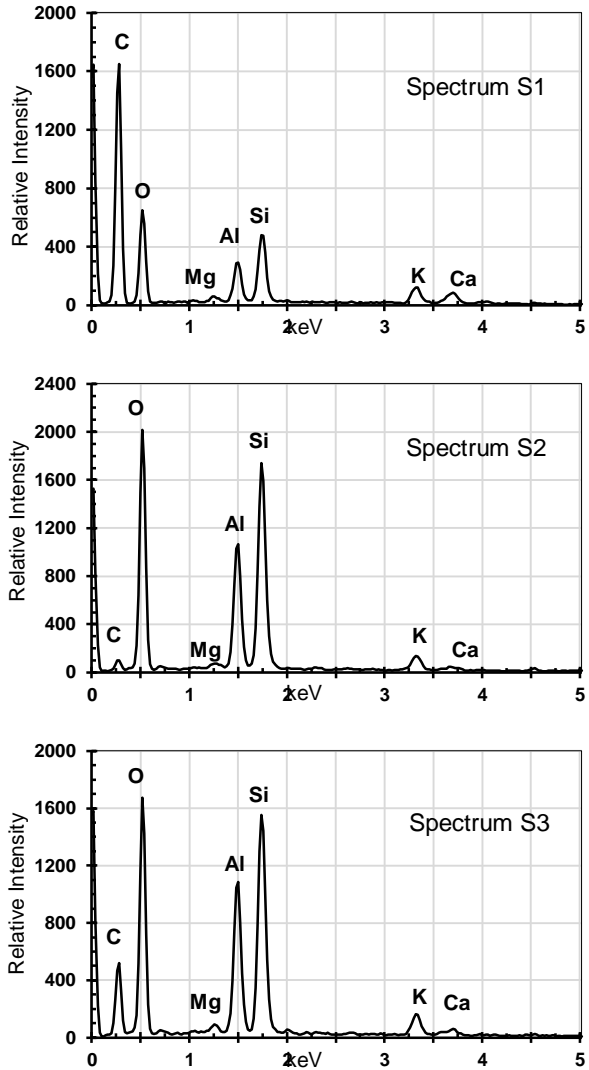

**Figure 5** Energy Dispersive X-ray analysis (EDX) results for the three points S1, S2 and S3 as displayed in Figure 4

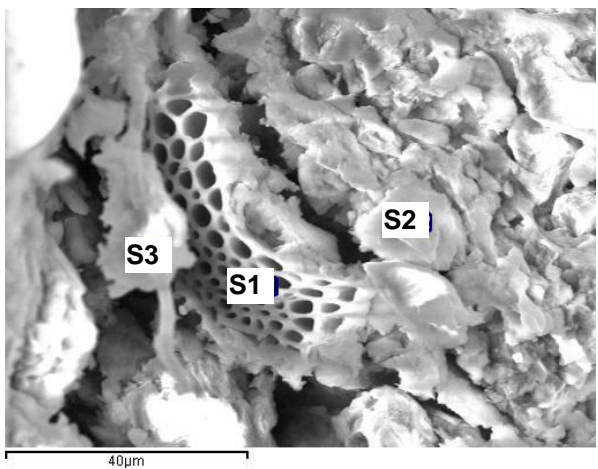

**Figure 4** Shows a highly carbonaceous partially degraded biological structure. Note the apparent fracture from the surrounding mineral matrix along the leading edge. Points S1, S2 and S3 correspond to elemental abundances determined by EDX as detailed in Table 2 and Figure 5

| Spec | S1 | S2 | S3 | Max | Min |
|---|---|---|---|---|---|
| C | 59.81 | 29.26 | 0.00 | 59.81 | 0.00 |
| N | 0.00 | 5.11 | 0.00 | 5.11 | 0.00 |
| O | 31.3 | 44.98 | 60.53 | 60.53 | 31.3 |
| Mg | 0.34 | 0.42 | 0.00 | 0.42 | 0.34 |
| Al | 2.11 | 6.68 | 12.15 | 12.15 | 2.11 |
| Si | 3.55 | 9.98 | 20.91 | 20.91 | 3.55 |
| P | 0.00 | 0.23 | 0.00 | 0.23 | 0.23 |
| S | 0.00 | 0.00 | 0.33 | 0.33 | 0.33 |
| K | 1.38 | 1.47 | 2.14 | 2.14 | 1.38 |
| Ca | 1.04 | 0.42 | 0.53 | 1.04 | 0.42 |
| Ti | 0.00 | 0.00 | 0.57 | 0.57 | 0.57 |
| Fe | 0.47 | 1.45 | 2.84 | 2.84 | 0.47 |

**Table 3** Elemental abundances as detected by energy X-ray analysis for points S1, S2 and S3 as displayed in Figure 4

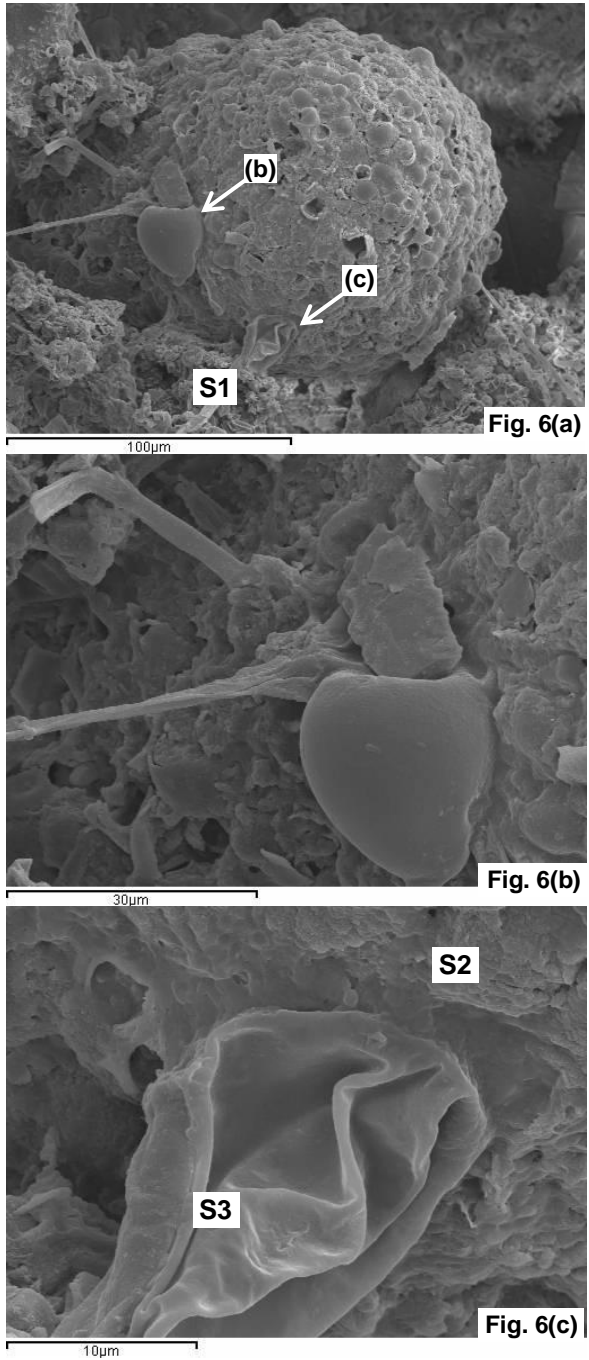

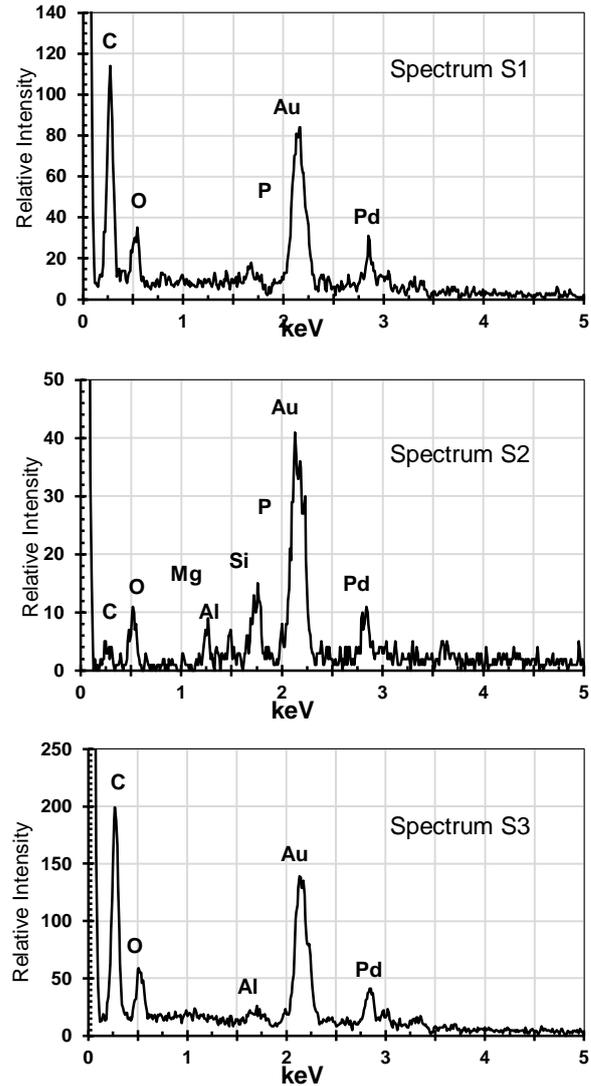

**Figure 7.** Energy X-ray analysis (EDX) results for the three points S1, S2 and S3 as displayed in Figure 6. Gold and Paladium readings are a result of sample coating.

**Figure 6 (a-c)** show images of a large (100 μm diameter) and very complex, thick-walled, carbon-rich (kerogenous) microfossil that we have tentatively identified as a hystrichosphere. Hystrichospheres are thought to be associated with the resting stages of a group of largely extinct marine dinoflagellate algae. The EDX data reveals a lack of detectable nitrogen and an anomalous C/O ratio (similar to bitumen) clearly establishing that this obviously biological form could not possibly represent a modern biological contaminant. The extremely well-preserved and very thin (2 μm diameter x 100 μm long) flagella are interpreted as indicating a low-gravity, low-pressure environment and rapid freeze-drying. In the vicinity of this form are highly carbonaceous embedded filaments of cyanobacteria ("blue-green algae"). Low nitrogen shows they are ancient fossils. Points S1, S2 and S3 relate to target corordinates of energy X-ray readings detailed in Figure 7 and Table 4. The sample was coated with a gold/palladium surface to facilitate high resolution image capture.

| Spec | S1 | S2 | S3 | Mean | SD | Max. |
|---|---|---|---|---|---|---|
| C | 67.94 | 0.00 | 65.73 | 44.56 | 38.60 | 67.94 |
| N | 0.00 | 0.00 | 0.00 | 0.00 | 0.00 | 0.00 |
| O | 26.21 | 39.82 | 24.80 | 30.28 | 8.29 | 39.82 |
| Na | 0.00 | 0.00 | 0.00 | 0.00 | 0.00 | 0.00 |
| Mg | 0.00 | 13.02 | 0.00 | 4.34 | 7.52 | 13.02 |
| Al | 0.00 | 10.35 | 0.00 | 3.45 | 5.98 | 10.35 |
| Si | 0.00 | 36.81 | 2.17 | 12.99 | 20.66 | 36.81 |
| P | 0.00 | 0.00 | 0.00 | 0.00 | 0.00 | 0.00 |
| S | 0.00 | 0.00 | 0.00 | 0.00 | 0.00 | 0.00 |
| K | 5.85 | 0.00 | 7.29 | 4.38 | 3.86 | 7.29 |
| Ca | 0.00 | 0.00 | 0.00 | 0.00 | 0.00 | 0.00 |
| Fe | 0.00 | 0.00 | 0.00 | 0.00 | 0.00 | 0.00 |

**Table 4.** Elemental abundances as detected by energy X-ray analysis for points S1, S2 and S3 as displayed in Figure 6. (Au and Pd subtracted).

Image recording is via a SONY video graphics printer or digital by processing image frames in a 16 bit framestore computer for output to hard drive.

Sample preparation involved the use of a flame sterilised wide-bore hypodermic needle to fracture the fragments before mounting on aluminium stubs. Further SEM studies were conducted at the NASA Marshall Space Flight Centre using uncoated freshly fractured surfaces that were examined using a Hitachi S-3700N Field Emission Scanning Electron Microscope with images gathered using both the Back-Scatter Electron and the Secondary Electron detectors.

Preliminary inspection of a few of the SEM images revealed the presence of a number of highly carbonaceous biological structures. Some of these were deeply integrated in the surrounding mineral matrix indicating they could not have been recent biological contaminants. EDX data indicates they are severely depleted in nitrogen (N < 0.5%). Hoover (2007) explored the use of Nitrogen levels and biogenic element ratios for distinguishing between modern and fossil microorganisms as a mechanism for recognizing recent biological contaminants in terrestrial rocks and meteorites. Detectable (2-18%) levels of Nitrogen were encountered in the hair and tissues from mummies from Peru (2 Kyr) and Egypt (5 Kyr) and the hair/tissues of Pleistocene Wooly Mammoths (40-32 Kyr). Comparative studies on fossilized insects in Miocene (8.4 Myr) Amber, Cretaceous Ammonites (100 Myr), Cambrian trilobites of Wheeler Shale of Utah (505 Myr) and filamentous cyanobacteria from Karelia (2.7 Gyr) showed nitrogen levels below the limit of detection with the FESEM EDS detector (N < 0.5%). This provides clear and convincing evidence that these obviously ancient remains of extinct marine algae found embedded in the Polonnaruwa meteorite are indigenous to the stones and not the result of post-arrival microbial contaminants.

## 6. Discussion

Results of initial studies on the Polonnaruwa stones were reported in a series of short bulletins in the *Journal of Cosmology* during the early part of 2013 (Wickramasinghe et al, 2013a, 2013b, 2013c, 2013d). A number of critical observations followed relating to the possible terrestrial origin of the stones as well as terrestrial contaminants being the most likely explanation for the observations. Critics also pointed to the unusual bulk composition of the stone (i.e. amorphous $SiO_2$), the presence of some crystalline phases of $SiO_2$ (quartz), the low density (<1g $cm^{-3}$) and the potential unreliability of eye witness reports. The hypothesis that the stones comprise of terrestrial fulgurite, the glassy product of tubes formed when quartzose sand, silica, or soil is heated by lightning strikes was suggested. However, the fulgurite hypothesis is readily disproved by the fact that the mineral lechatelierite (which comprises fulgurites) has an average density of 2.57 g $cm^{-3}$, whereas the density of the Polonnaruwa meteorites is <1 g $cm^{-3}$.

Furthermore, none of the Polonnaruwa meteorites studied are in the form of hollow tubes, which is a well-known common characteristic of fulgurites. Additionally, due to the method in which they form, fulgurites typically have to be dug out of the soil. In contrast, the Polonnaruwa stones were found lying on the surface of undisturbed sandy soil in a rice field. The eyewitnesses reported that there was no lightning activity in this region on the night of Dec. 29, 2012. To have produced such a large number of stones as have been found in a small field would have required cloud to ground lightning strikes many orders of magnitude greater than ever before reported. In addition, the temperature at which sand must be heated by lightning to melt and form a fulgurite (1770 C) would certainly have vaporized and burned all carbon-rich biological (cyanobacterial filaments, hystrichospheres, etc.) and melted and thus destroyed the delicately marked silica frustules of the diatoms.

Subsequent suggestions that an "unusually porous" fulgurite may have floated in on water ignores the fact that many of these low-density (<1 g $cm^{-3}$) stones were recovered between Dec. 29, 2012 and Jan. 29, 2013 and found to exhibit similar mineralogy, morphology, physical and microbiological properties. None of these original stones studied at Cardiff University or at the NASA Marshall Space Flight Center were observed to be hollow tubes and none were found to be composed of the mineral lechatelierite. The "floating on water" hypothesis is also inconsistent with FTIR data obtained using diamond ATR instrumentation that show a conspicuous absence of a characteristic water peak (Wickramasinghe et al, 2013c). This observation alone appears to rule out the likelihood of aqueous-based contaminants being responsible for the presence of observed diatom frustules in either a non-terrestrial fragment or a formerly sterile terrestrial fulgurite.

Finally, the results of triple-oxygen isotope analysis presented here give a $\Delta^{17}O$ value of - 0.335. This result, together with previous studies on the oxygen isotope composition of fulgurites (Robert and Javoy, 1992) that find no marked isotope exchange with atmospheric molecular oxygen during the formation process provides a strong case against the fulgurite conjecture.

We are aware that a large number of unrelated stones have been submitted for analysis, and we have no direct information or knowledge regarding the nature, source, origin or provenance of the stones our critics have examined. Based upon the physical, chemical, isotopic, mineralogical and microbiological studies that we have performed, we are certain that the Polonnaruwa stones submitted to us for examination by the Medical Research Institute in Sri Lanka and the stones that we recovered from the rice field (7 º 52' 59.5" N; 81 º 9' 15.7" E) on Jan. 29, 2013 were not fulgurites, scoria, lava, pumice or the product of industrial processes.

The presence of a number of carbonaceous biological structures exhibiting severe nitrogen depletion is highly indicative of ancient fossilised biological remains. Some of these were deeply integrated in the surrounding mineral matrix suggesting they could not have been recent terrestrial contaminants. The presence of these structures may have further implications since diatoms are known to be unable to break the strong triple bond of the inorganic di-nitrogen ($N_2$) molecule and convert it into the organic nitrogen. Nitrogen in molecules such as $NO_2$ or $NH_3$ can be readily used for the construction of amino acids, proteins, DNA, RNA and other life-critical biomolecules. The function of nitrogen-fixation is carried out by the nitrogenous enzyme that is housed in structures known as heterocysts of several species of several genera (Nostoc Tolypothrix, etc.) of heterocystous cyanobacteria. Consequently, cyanobacteria are considered a crucial component of a cometary ecosystem.

We conclude that the oxygen isotope data show P159/001-03 and P/159001-04 are unequivocally meteorites, almost certainly fragments originating from the fireball-causing bolide. The most likely origin of this low density meteorite with delicate structures, some highly carbonaceous, is a comet (Wickramasinghe et al., 2013b). The presence of fossilized biological structures provides compelling evidence in support of the theory of cometary panspermia first proposed over thirty years ago (Hoyle and Wickramainghe, 1981, 1982, 2000).

**Acknowledgement**

We are grateful to Professor Andreas Pack of the University of Göttingen, Germany for his assistance with the Triple Oxygen Isotope analysis.

**REFERENCES**

Cerling, T. E., and Quade, J., 1993, Stable Carbon and Oxygen Isotopes in Soil Carbonates, In P.K. Swart, K.C. Lohmann, J.A.McKenzie, and S.M.Savin, eds., *Climate Change in Continental Isotope Records*, pp217-231.

Clayton, R. N., and Mayeda, T. K., 1999, Oxygen Isotope Studies of carbonaceous chondrites, *Geochimica et Cosmochimica Acta*, **Vol 63,** No 13/14 pp2089-2104.

Franchi, I.A., Wright, I.P., Sexton, A.S., and Pillinger, C.T. 1999, The oxygen isotope composition of Earth and Mars, *Meteoritics & Planetary Science* **34**, pp657-661.

Gehler, A., Tütken, T., and Pack, A., 2011, Triple oxygen isotope analysis of bioapatite as tracer for diagenetic alteration of bones and teeth, *Palaeogeography, Palaeoclimatology, Palaeoecology*, **Vol 310**, 1-2 pp 84-91.

Greenwood R.C., Franchi I.A., Jambon A. and Buchanan P.C. (2005) Widespread magma oceans on asteroidal bodies in the early solar system. *Nature* **435**, 916-918.

Hoover, R. B., 2007, Ratios of Biogenic Elements for Distinguishing Recent from Fossil Microorganisms. *Instruments, Methods, and Missions for Astrobiology X, Proc.* SPIE **66940D**.

Hoyle, F. and Wickramasinghe, C., "Proofs that Life is Cosmic", Memoirs of the Institute of Fundamental Studies, Sri Lanka, No. 1, Dec 1982.

Hoyle, F. and Wickramasinghe, N.C., 1981. "Comets - a vehicle for panspermia", in Comets and the Origin of Life, ed. C. Ponnamperuma, D. Reidel Publishing Co. Dordrecht.

Hoyle, F. and Wickramasinghe, N.C., 2000. "Astronomical Origins of Life – Steps towards panspermia" (Kluwer Academic Publishers, Dordrecht).

Ikeda Y. An overview of the research consortium, "Antarctic carbonaceous chondrites with CI affinities, Yamato-86720, Yamato-82162, and Belgica-7904." Proc. NIPR Symp. Antarct. Meteorites, 5, 49–73, (1992).

Ivanova M. A., Taylor L. A., Clayton R. N., Mayeda T. K., Nazarov M. A., Brandstaetter F., and Kurat G., Dhofar 225 vs. the CM clan: Metamorphosed or new type of carbonaceous chondrite? (abstract). In Lunar and Planetary Science XXXIII, Abstract #1437. Lunar and Planetary Institute, Houston, (2002).

Pack, A., 2008, Fractionation of refractory lithophile elements in bulk chondrites and chondrite components. *Lunar and Planetary Science Conference, Houston*.

Robert, F., and Javoy, M., 1992, Oxygen Isotope Compositions of Fulgarites, *Meteoritics* **27** (1992): 281.

Tameoka, K. 1990. Mineralogy and Petrology of Belgica-7904: A New Kind of Carbonaceous Chondrite from Antarctica. Proc. NIPR Symp. Antarct. Meteorites. 3, 40-54, 1990.

Wickramasinghe, N.C., Wallis, J., Wallis, D.H., Samaranayake, A., 2013a. Fossil diatoms in a new carbonaceous meteorite, *Journal of Cosmology*, **21**, 37.

Wickramasinghe, N.C., Wallis, J., Wallis, D.H., Wallis M.K., Al-Mufti, S., Wickramasinghe, J.T., Samaranayake, A. and Wickramarathne, K., 2013b. On the cometary origin of the Polonnaruwa meteorite, *Journal of Cosmology*, **21**, 38.

Wickramasinghe, N.C., Wallis, J., Miyake, N., Wallis D.H., Samaranayake, A., Wickramarathne, K., Hoover, R., and Wallis M.K., 2013c. Authenticity of the life-bearing Polonnaruwa meteorite, *Journal of Cosmology*, **21**, 39.

Wickramasinghe, N.C., Samaranayake, A., Wickramarathne, K., Wallis, D.H., Wallis M.K., Miyake, N., Coulson, S.J., Hoover, R., Gibson, C.H., and Wallis, J.H., 2013d, Living Diatoms in the Polonnaruwa meteorite – Possible link to red and yellow rain, *Journal of Cosmology*, **21**, 40.